\documentstyle[12pt,a41,epsfig,portland,cite]{article}

\newcommand{\Mvec}{\mbox{\rm\bf M}}

\newcommand{\beq}{\begin{equation}}
\newcommand{\eeq}{\end{equation}}
\newcommand{\bea}{\begin{eqnarray}}
\newcommand{\eea}{\end{eqnarray}}

\newcommand{\gsim}{\raisebox{-0.07cm}{$\, \stackrel{>}{{\scriptstyle
\sim}}\, $}}

\newcommand\GeV{\,\mbox{GeV}}

\newcounter{lin}

\begin{document}
\begin{titlepage}

\begin{flushleft}
DESY 03--204 \hfill {\tt hep-ph/0404034} 
\\
SFB-CPP/04--02\\
December 2003                         \\
\end{flushleft}

\vspace{3cm}
\begin{center}
{\LARGE\bf Mellin Representation for the Heavy Flavor} 

\vspace{3mm}
{\LARGE\bf Contributions to Deep Inelastic}

\vspace{3mm}
{\LARGE\bf Structure Functions} 

\vspace{4cm}
{\large Sergey I. Alekhin\footnote{On leave of absence from IHEP, 
RU--142281 Protvino, Russia} and Johannes Bl\"umlein}

\vspace{2cm}
{\large\it 
DESY--Zeuthen}\\

\vspace{3mm}
{\large\it  Platanenallee 6, D--15735 Zeuthen, Germany}\\

\vspace{3cm}
\end{center}
\begin{abstract}
\noindent
We derive semi--analytic expressions for the analytic continuation of the
Mellin transforms of the heavy flavor QCD coefficient functions for neutral 
current deep inelastic scattering in leading and next-to-leading order to
complex values of the Mellin variable $N$. These representations are used
in Mellin--space QCD evolution programs to provide fast evaluations of the
heavy flavor contributions to the structure functions $F_2(x,Q^2), F_L(x,Q^2)$ 
and $g_1(x,Q^2)$. 
\end{abstract}

\end{titlepage}

\newpage
\sloppy
\section{Introduction}
\label{sec:intro}

\vspace{2mm}
\noindent
A major goal in investigating the scaling violations of the structure  
functions as measured in deeply inelastic scattering is to determine the
QCD scale parameter $\Lambda_{\rm QCD}$, or equivalently the strong
coupling constant $\alpha_s(\hat{Q}^2)$, at a typical reference scale
$\hat{Q}^2$. The QCD renormalization group equations for mass
factorization 
at space--like momentum transfer are solved referring to either $x$--space 
\cite{XSPAC}\footnote{A series of these codes uses orthogonal polynomials
to diagonalize the evolution equations~\cite{ORTHPOL}.} or Mellin--space
\cite{MSPACE,BB} representations.~\footnote{For numerical comparisons of
different evolution codes see \cite{COMP}.}
The formulation of the evolution
equations in Mellin space has the advantage that only {\sf ordinary}
differential equations, unlike integro-differential equations in
$x$-space, have to be solved. Moreover, to maintain factorization--scheme
invariance the evolution equations have to be consistently expanded in the
strong coupling constant, which allows an analytic integration of the
differential equations, i.e. a complete analytic solution. The only
numerical element consists in the inverse Mellin transform of the solution
to $x$--space via a contour integral around the singularities of the
solution in the complex $N$--plane. The solution in fixed order
perturbation theory has isolated poles on the real axis bounded by some
positive number from above. Resummations of small--$x$ contributions can
be easily implemented in this approach in a {\sf consistent}
way~\cite{SX}. 
In the QCD fit only the parameters of the
non--perturbative input distributions and $\Lambda_{\rm QCD}$ have to 
be varied and the evolution kernels consisting out of the
anomalous dimensions and the Wilson coefficients to the respective order
in the coupling constant need not to be recalculated  but
are kept fixed as (tabulated) $Q^2$--dependent arrays along the contour
used for the numerical Mellin inversion. This makes QCD fits using
Mellin--space programs particularly fast.

The analytic solution of the evolution equations in Mellin space allows
to study factorization--scheme invariant evolution equations of different
kind~\cite{SI,BB}. These  equations describe the evolution of {\sf
physical observables}. In the non--singlet case a flavor non--singlet
combination of structure functions forms the input distribution, which can
be measured at a starting scale $Q^2_0$. The usual flavor singlet
evolution can be mapped into the evolution of two structure functions as
e.g. $F_2(x,Q^2)$ and $F_L(x,Q^2)$, $F_2(x,Q^2)$ and $\partial
F_2(x,Q^2)/\partial \ln(Q^2)$, or $g_1(x,Q^2)$ and $\partial
g_1(x,Q^2)/\partial \ln(Q^2)$, \cite{BB}.
Here the non--perturbative input 
distributions are formed by the measured structure functions at a scale
$Q^2_0$, and have therefore not to be determined in the QCD--analysis.
The
evolution kernels  in $x$--space are complicated multiple Mellin
convolutions of Wilson coefficients, splitting functions $P_{ij}^{(k)}(z)$
and of some Mellin--inverse splitting functions and Wilson coefficients in
general,~see~\cite{BRN1}. These physical evolution kernels can be easily 
calculated in analytic form in Mellin--$N$ space unlike in $x$--space,
where, as is well known, already the Mellin--inverse of leading order 
splitting
functions takes a complicated form~\cite{BK} and the convolutions to be
performed were never done analytically.
Due to the sizeable contribution of charm to the structure functions
$F_2(x,Q^2)$ and $F_L(x,Q^2)$ in the kinematic regime of HERA ($x \gsim
10^{-4}, Q^2 > 5 \GeV^2$) an exact treatment requires the account for   
heavy quark effects on $F_2(x,Q^2)$ and $F_L(x,Q^2)$ which is easiest
described in analytic form in Mellin--$N$ space. The standard QCD 
analysis and fits using scheme--invariant evolution provide different 
methods to determine the strong coupling constant $\alpha_s(M_Z^2)$. 
In 
the upcoming high--precision analyses, which aim on measurement errors 
of $O(1\%)$, it is of particular importance to diminish all theoretical 
and conceptual uncertainties. For this purpose the use of both these 
methods will be essential.

So far no specific representation of the heavy flavor Wilson coefficients
in Mellin--$N$ space, in the way used for those of the light flavors, were
derived. It is the aim of the present paper to present these
parameterizations in terms of a semi-analytic approach. The representation
is based on meromorphic functions in the Mellin variable $N$. The
$\xi = Q^2/m^2$ dependence is given by a series of numerical expansion
coefficients until a desired accuracy is reached. Interpolations
in the variable $\xi$ are used along the inversion contour to keep the
analysis as fast as possible. The present parameterizations enable to
perform scheme--invariant evolutions including the effect of heavy flavor
coefficient functions. 

The paper is organized as follows. In section~2 we summarize the basic
notations to describe the neutral--current heavy flavor contributions 
to deep--inelastic scattering up to next--to--leading
order~\cite{LOUNP,SVZ,LOPOL,HFNLO,RSN}. In section~3 we present the
parameterization of the different contributions to the QCD Wilson
coefficients for heavy flavor production for complex values of
Mellin--$N$ and apply these representations to calculate the heavy flavor
structure functions using sample densities for the light partons,
including the study of their accuracy. Section~4 contains the conclusions. 
\section{The Heavy Flavor Structure Functions}
\label{sec:cross}

\vspace{2mm}
\noindent
The Wilson--coefficients for heavy flavor production in deeply inelastic
scattering depend on two different scales, which can be represented
choosing different kinematic variables. Here we follow, for convenience,
the notation of Ref.~\cite{RSN} and use the variables $\xi$ and $\eta$~: 
\begin{eqnarray}
\xi = \frac{Q^2}{m^2},~~~~~~~~~~~~~~~~~~\eta = \frac{s}{4 m^2} - 1 \geq
0~.
\end{eqnarray}
$s$ denotes the cms energy squared of the heavy quark system, $Q^2=-q^2$
the
four--momentum transfer squared, and $m$ the heavy quark mass. The
momentum fraction $z$ carried by the struck parton in the nucleon is
\begin{eqnarray}
\label{eqZ}
z = \frac{Q^2}{Q^2+s} = \frac{\xi/4}{1+\eta +\xi/4}~,~~~~~~~~~~~~~ 
z~\epsilon~\left[x,~
\frac{Q^2}{Q^2 + 4 m^2}\right]~,
\end{eqnarray}
and the cms velocity of the heavy quarks is
\begin{eqnarray}
v  = \sqrt{1- \frac{4 m^2}{Q^2} \frac{z}{1-z}} =
\sqrt{\frac{\eta}{1+\eta}}~.
\end{eqnarray}
The heavy flavor structure functions are given in leading (LO) and
next-to-leading order (NLO) by 
\begin{eqnarray}
\label{eqSF}
F_k(x,Q^2,m^2) &=& F_k^{\rm LO}(x,Q^2,m^2) + F_k^{\rm NLO}(x,Q^2,m^2)~,\\
G_k(x,Q^2,m^2) &=& G_k^{\rm LO}(x,Q^2,m^2)~,
\end{eqnarray}
with $k = 2, L$ for the unpolarized structure functions and $k=1$ for the
polarized structure functions. In the latter case the NLO corrections are
not yet calculated. The heavy flavor contributions to the structure
function $g_2(x,Q^2)$ are obtained at leading twist from $G_1(x,Q^2,m^2)$
via the Wandzura--Wilczek relation~\cite{BNR2}. The leading order
contributions are
\begin{eqnarray}
\label{HkLO} 
H_k^{\rm LO}(x,Q^2,m^2) =
\frac{Q^2}{\pi m^2} a_s(\mu^2)  e_Q^2 \int_x^{z_{\rm max}}~\frac{dz}{z}~
h_g\left(\frac{x}{z}, \mu^2\right) c_{g,H_k}^{(0)}(\eta,\xi)~,
\end{eqnarray}
with $H = F, G,~h =f, g$, $f_g(z,\mu^2) = G(z,\mu^2), g_g(z,\mu^2) = \Delta
G(z,\mu^2)$ the unpolarized and polarized gluon distribution functions,
resp., and an implicit $z$ dependence of the Wilson--coefficients
$c_{g,H_k}^{(0)}(\eta,\xi)$. $e_Q$ denotes the electric charge of the
heavy
quark, $a_s = \alpha_s/(4\pi)$ the strong coupling constant, and $\mu^2$ 
the factorization scale. The
LO Wilson coefficients
are~\cite{LOUNP, SVZ, LOPOL}~:
\begin{eqnarray}
c_{g,F_L}^{(0)}(\eta,\xi) &=& \frac{\pi}{2} T_f \frac{\xi}{(1+\eta +
\xi/4)^3}\left\{ 2\left[\eta(1+\eta)\right]^{1/2} - L(\eta)\right\}~,\\
c_{g,F_T}^{(0)}(\eta,\xi) &=& 
\frac{\pi}{2} T_f \frac{1}{(1+\eta + \xi/4)^3} 
\left\{
-2 \left[(1+\eta-\xi/4)^2+1+\eta\right]
\left(\frac{\eta}{1+\eta}\right)^{1/2}
\right.\nonumber\\  && 
\left. + 
\left[2(1+\eta)^2+\frac{\xi^2}{8} + 1+2\eta\right] 
L(\eta) \right\}~,\\
c_{g,F_2}(\eta,\xi) &=& c_{g,F_L}(\eta,\xi) + c_{g,F_T}(\eta,\xi)\\
c_{g,g_1}^{(0)}(\eta,\xi) &=& \frac{4\pi}{\xi} T_f \frac{1}{1+\eta+\xi/4}
\left\{\sqrt{\frac{\eta}{1+\eta}}\left[3(1+\eta) - \frac{\xi}{4}\right]
- \left(1 + \eta - \frac{\xi}{4} \right) L(\eta)\right\}~, 
\end{eqnarray}
with $T_f = 1/2$ in $SU(N)$ and 
\begin{eqnarray}
L(\eta) = \ln \left[\frac{(1+\eta)^{1/2} + \eta^{1/2}} 
                         {(1+\eta)^{1/2} - \eta^{1/2}}\right]~.
\end{eqnarray}

The NLO contributions $F_k^{\rm NLO}(x,Q^2,m^2)$ are given by~: 
\begin{eqnarray}
\label{FkNLO}
F_k^{\rm NLO}(x,Q^2,m^2) &=& \frac{Q^2}{\pi m^2} \alpha_s^2(\mu^2)
\int_x^{z_{\rm max}}~\frac{dz}{z}~\Biggl\{ e_Q^2 f_g\left(\frac{x}{z},
\mu^2
\right)\left[c_{k,g}^{(1)}(\xi,\eta) + \overline{c}_{k,g}^{(1)}(\xi,\eta)
\ln\left(\frac{\mu^2}{m^2}\right) \right] \nonumber\\ 
&& \hspace{2.0cm} + \sum_{i = q, \overline{q}}^3 \Biggl\{ 
e_Q^2 f_i\left(\frac{x}{z}, \mu^2\right) \left[c_{k,i}^{(1)}(\xi,\eta) +
\overline{c}_{k,i}^{(1)}(\xi,\eta)
\ln\left(\frac{\mu^2}{m^2}\right) \right] \nonumber\\
&& \hspace{3.15cm} + e_i^2 f_i\left(\frac{x}{z},
\mu^2\right) \left[d_{k,i}^{(1)}(\xi,\eta) + 
\overline{d}_{k,i}^{(1)}(\xi,\eta)\right]~,
\end{eqnarray}
with $\overline{d}_{L,q}^{(1)}(\xi,\eta)=0$.\footnote{These functions 
emerge in case one would like to include  photo-production as 
well.} 
In 
\cite{RSN} the functions $c_{k,j}^{(1)}, d_{k,j}^{(1)}$ and
$\overline{c}_{k,j}^{(1)}, \overline{d}_{k,j}^{(1)}$ are parameterized
in terms of analytic expressions for the threshold and asymptotic region
and a tabulated numerical term interpolating between both.

\section{Parameterization}
\label{sec:para}

\vspace{2mm}
\noindent
The parameterization of the Mellin transform for the Wilson coefficients 
in section~\ref{sec:cross} shall be carried out for complex values of the
Mellin variable $N$.
Complete analytic results for the  Mellin transforms of the heavy
flavor coefficient functions are difficult to obtain even in lowest order
in QCD\footnote{As shown in Ref.~\cite{SVZ} for the unpolarized
leading order coefficient functions analytic results for the {Mellin}
transform can be calculated for {\it integer} $N$ are given in a form
which does not allow analytic continuation.}. A very efficient
method in representing higher order functions to high precision,
cf.~\cite{MINIM1}, consists in
the {\tt MINIMAX}--method, see also~\cite{MINIMAX}. This polynomial 
representation uses Chebyshev--polynomials in the approximation. If 
compared to the Taylor--expansion of a function the coefficients are
adapted in such a way that the convergence is significantly better 
comparing
polynomials of the same degree. This method has been
also successfully being used in representing analytic
continuations~\cite{ANCONT} of Mellin transforms occurring in massless higher
order calculations~\cite{HO1} before.

The heavy quark structure function (\ref{eqSF}) are Mellin convolutions
\begin{eqnarray}
\label{eqSF1}
F(x) = (A \otimes C)(x) = \int_0^1 dx_1 \int_0^1 dx_2 A(x_1) C(x_2)
\delta(x - x_1 x_2)~,
\end{eqnarray}
where the parton densities have support $x_1~\epsilon~[0,1]$
and the coefficient function $x_2~\epsilon~[0,\xi/(\xi+4)]$. (\ref{eqSF1})
can be diagonalized by the Mellin transform 
\begin{eqnarray}
\label{eqMEL1}
\Mvec[F(x)](N) &=& \int_0^1 dx x^{N-1} F(x) \\
\Mvec[F(x)](N) &=& \Mvec[A(x_1)](N) \cdot \Mvec[C(x_2)](N)~,
\end{eqnarray}
accounting for the support in $C(x_2)$. The Wilson--coefficients
$C(z,\xi)$ are 
now parameterized as  polynomials in $z$ in the range $[0,\rho]$ with
$\rho =\xi/(\xi+4)$ at
fixed values of $\xi$ using the {\tt MINIMAX}--method to obtain optimal
convergence. Sometimes it is useful to study the function  
\begin{eqnarray}
\label{eqCO1}
C^{\tt MINIMAX}(z,\xi) (\rho-z)^\kappa = \sum_{k=0}^{K} a_k(\rho) z^k
\end{eqnarray}
instead of $C(z,\xi)$ to improve the convergence.

The Mellin transform of $C(z,\xi)$ is then obtained by a
Riemann--Liouville fractional integral~\cite{BAT}
\begin{eqnarray}
\label{eqMEL2}
\Mvec[C(z,\xi)](N) = \sum_{k=0}^{K} a_k(\rho) \rho^{N+k-\kappa}
B(N+k,1-\kappa)~.
\end{eqnarray}
Here, $B(a,b)$ denotes the Euler Beta-function.
As evident, (\ref{eqMEL2}) defines a meromorphic function in $N$ with
poles on the real axis bounded from above. For $\kappa = 0$,~~~$B(N+k,1)
= 1/(N+k)$.
In all representations below we refer to 15 expansion coefficients
$a_k(\rho)$, i.e. $K = 14$. In a few cases a representation with $\kappa
\neq 0$ is chosen. We calculate the coefficients $a_k(\rho)$ in
300 logarithmic steps in $\xi~\epsilon~[0.4,~10^4]$ and use 
numerical interpolation.

In Figures~1a--c we show the LO coefficient functions $c_{g,F_L}^{(0)},
c_{g,F_2}^{(0)}$ and $c_{g,g_1}^{(0)}$ and the absolute accuracy of their
representation by the corresponding {\tt MINIMAX}--polynomial (\ref{eqCO1}).
Figures~1d, e show examples for the {\tt MINIMAX}--representation for
NLO contributions to the heavy flavor Wilson coefficients.
In Figures~1a--e below the accuracy of the representation $\Delta
C(\xi,\eta)$ is evaluated as
\begin{eqnarray}
\label{eqACCc}
\Delta C_i(\xi,\eta) = |C_i^{\tt MINIMAX}(\xi,\eta) - C_i(\xi,\eta)|~.
\end{eqnarray}
Similar results as shown in Figures~1a--e are obtained for the other
Wilson coefficients.

Table~1 gives a complete survey on the absolute accuracy of the
representations of the Wilson coefficients in (\ref{HkLO},\ref{FkNLO}). 
In many of the cases very small maximal absolute errors, i.e. of $O(10^{-5})$ 
and smaller, of the Wilson coefficients were obtained in the whole $\xi$--range.  
In some cases the errors may reach $\sim 6 \cdot 10^{-2}$, however. 
Our main goal is to obtain sufficiently accurate representations for the heavy flavor
structure functions.
\newpage  
\begin{center}

\vspace*{-7mm}
\mbox{\epsfig{file=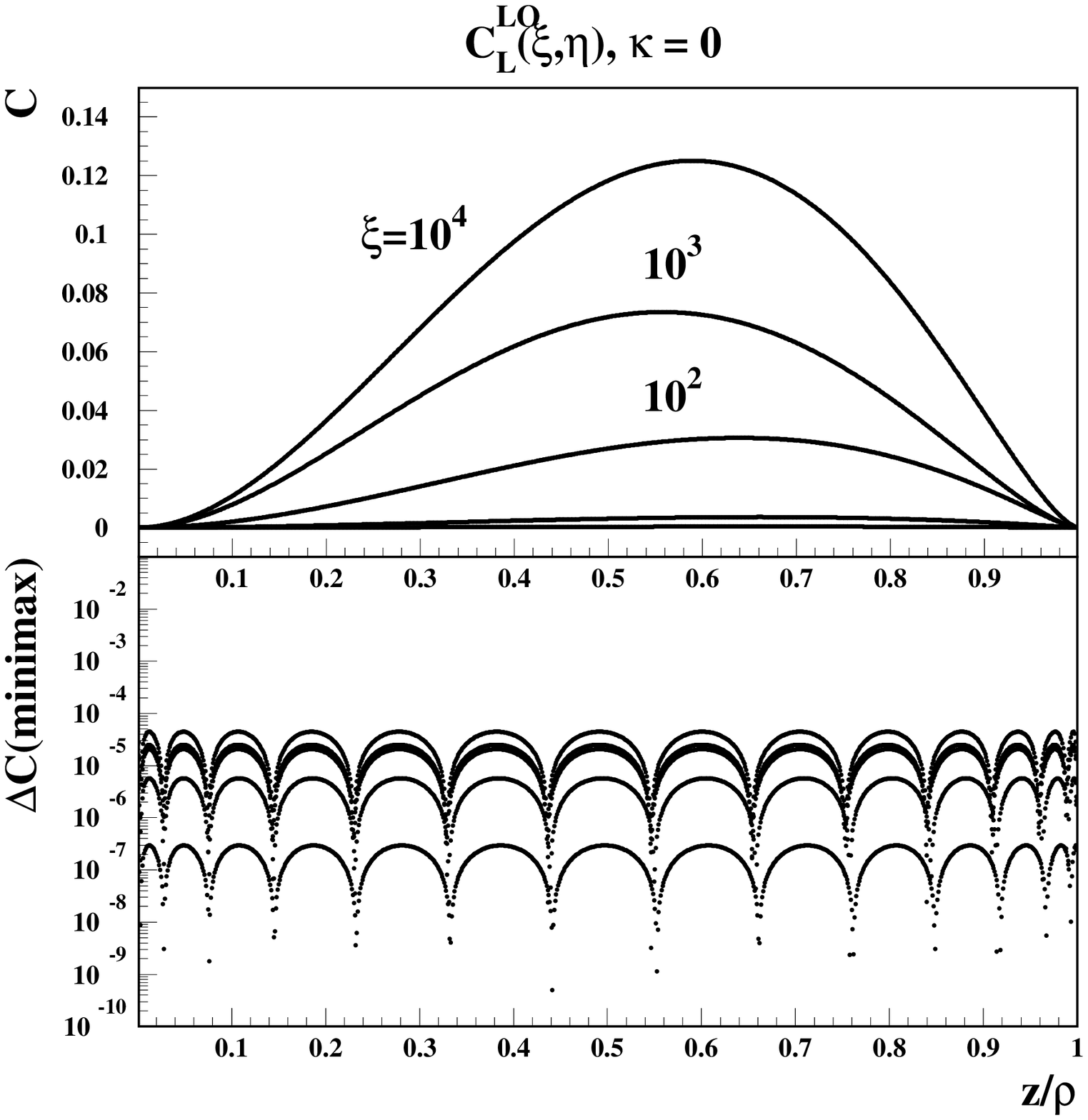,height=10cm,width=10cm}}

\vspace{2mm}  
\noindent
\small
\end{center}
{\sf
Figure~1a:~LO Wilson coefficient $c_{F_L,g}^{(0)}$ as a function
of $z/\rho$ and $\xi= 1, 10, 10,^2, 10^3, 10^4$ and the modulus of the error of the
polynomial 
representation, $\Delta C$.}
\normalsize 
\begin{center}

\mbox{\epsfig{file=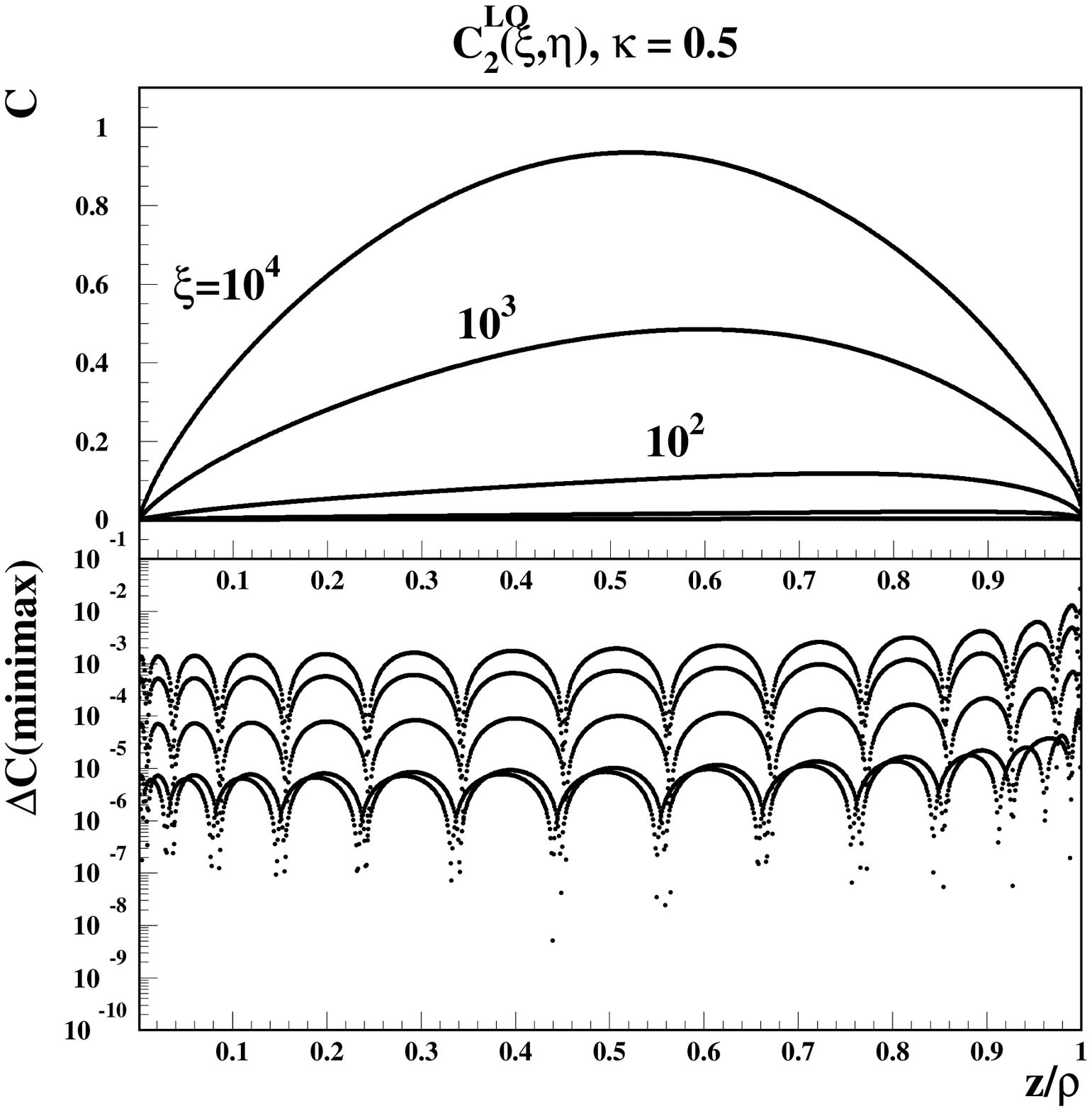,height=10cm,width=10cm}}

\vspace{2mm}
\noindent
\small
\end{center}
{\sf
Figure~1b:~LO Wilson coefficient $c_{F_2,g}^{(0)}$. The {\tt
MINIMAX}-polynomial was
determined choosing $\kappa = 0.5$ in (\ref{eqCO1}). All other conditions
are as in Figure~1a.}
\normalsize
\newpage
\begin{center}

\mbox{\epsfig{file=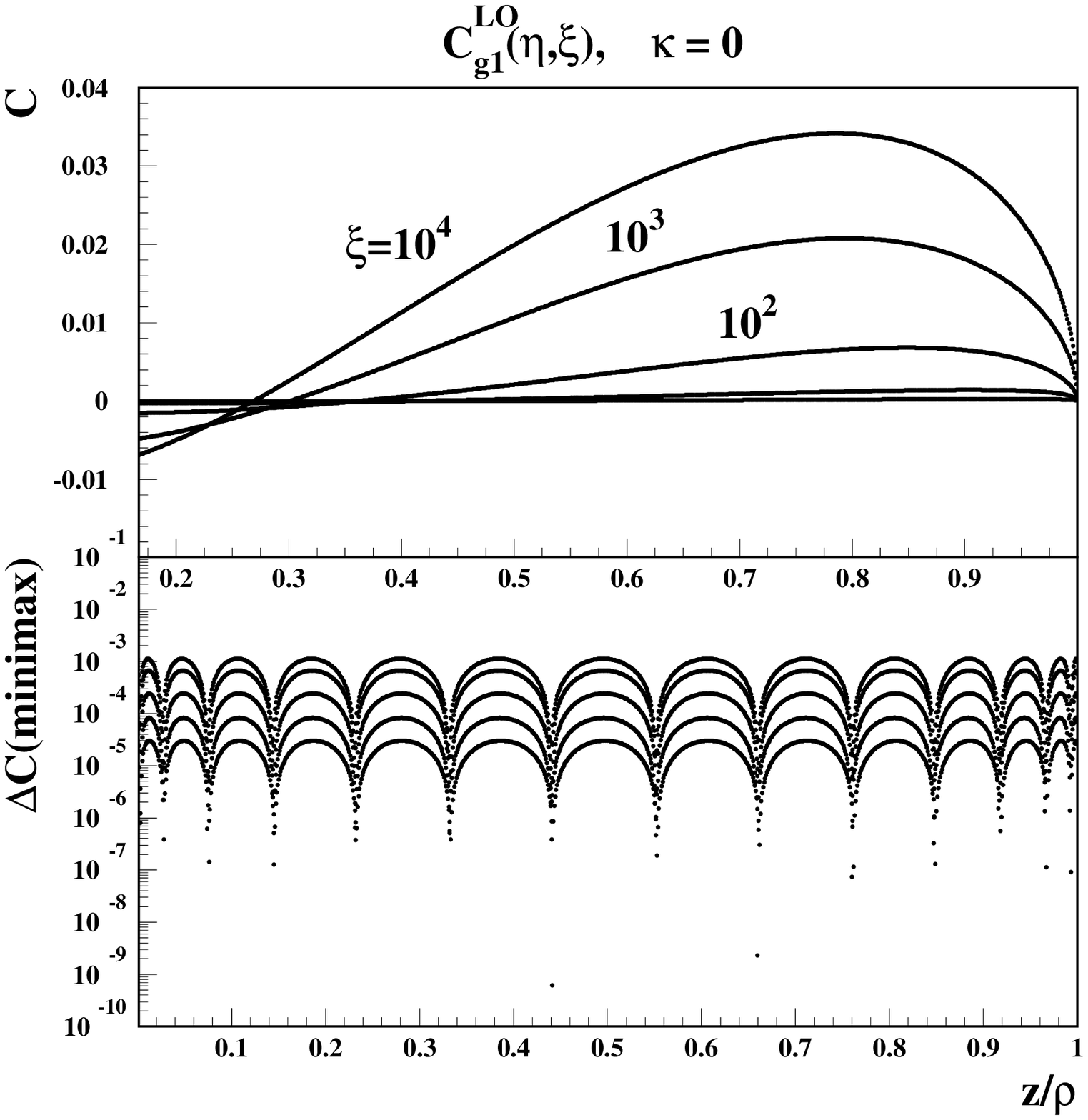,height=10cm,width=10cm}}

\vspace{2mm}
\noindent
\small
\end{center}
\begin{center}
{\sf
Figure~1c:~LO Wilson coefficient $c_{g_1,g}^{(0)}$. 
All other conditions
are as in Figure~1a.}
\end{center}
\normalsize
\begin{center}

\mbox{\epsfig{file=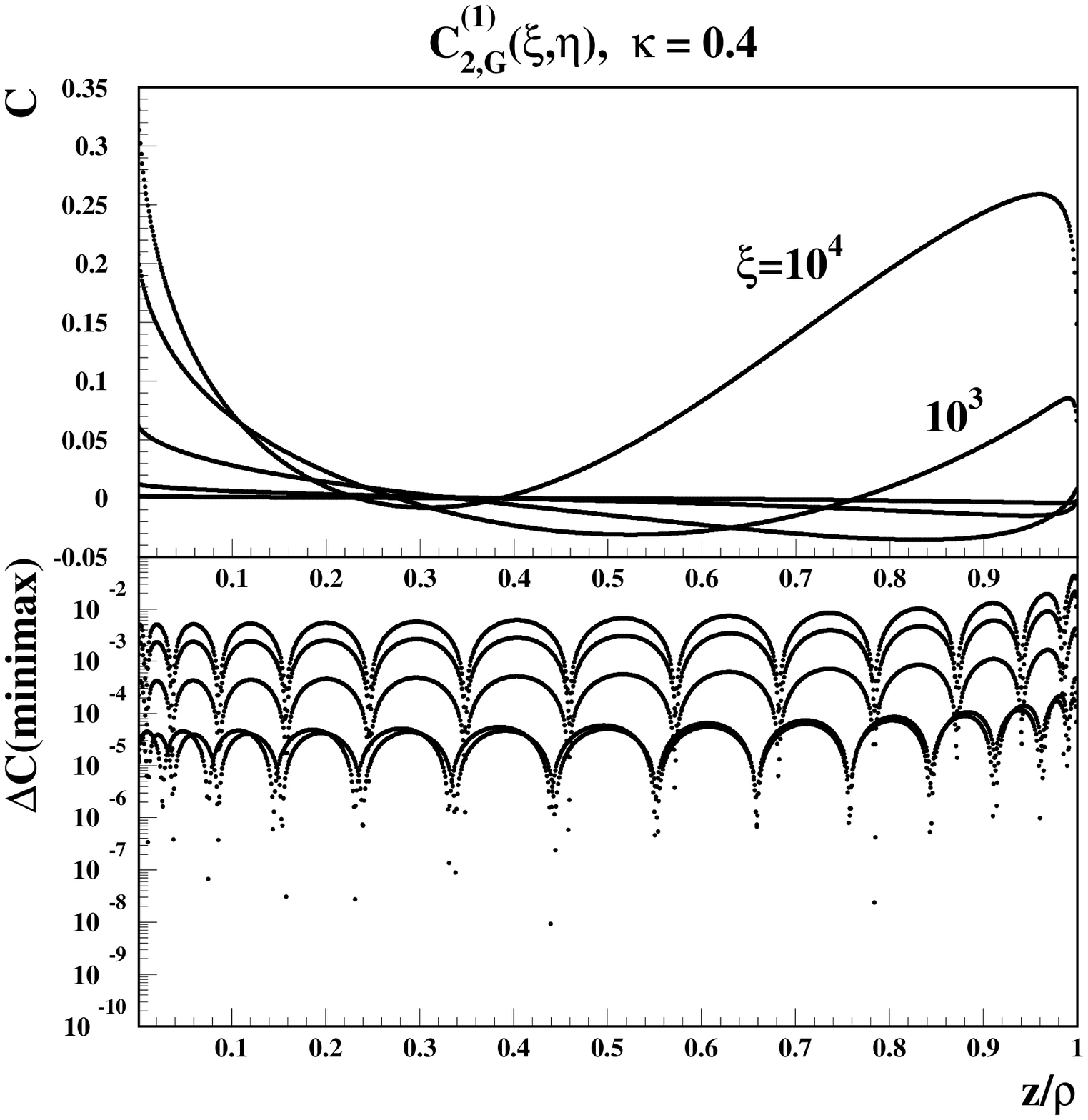,height=10cm,width=10cm}}

\vspace{2mm}
\noindent
\small
\end{center}
{\sf
Figure~1d:~Contribution to the NLO Wilson coefficient $c_{F_2,g}^{(1)}$.
The {\tt
MINIMAX}-polynomial was determined choosing $\kappa = 0.4$ in
(\ref{eqCO1}). All other conditions are as in Figure~1a.}
\normalsize
\newpage

\vspace*{-5mm}
\begin{center}

\mbox{\epsfig{file=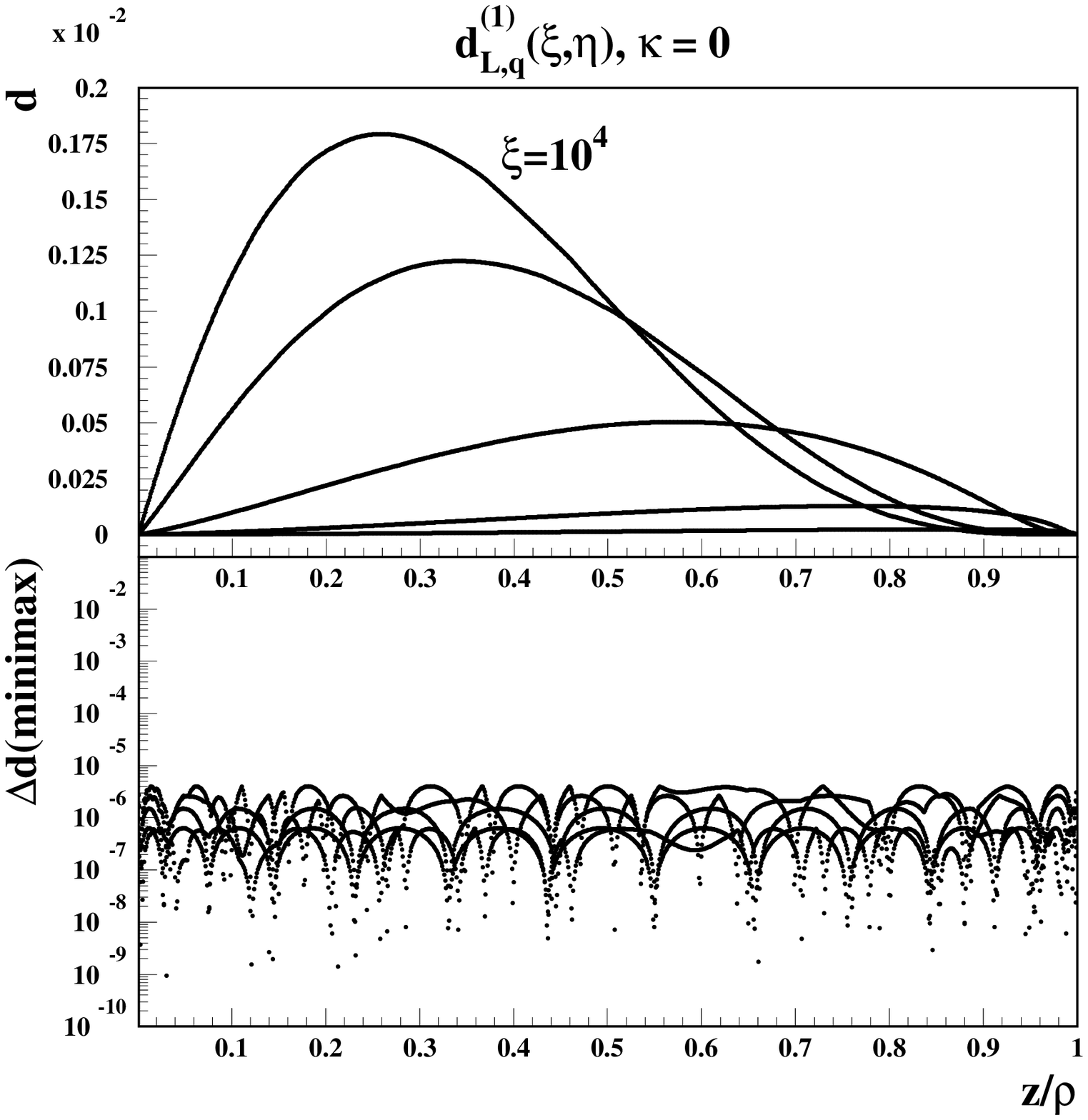,height=10cm,width=10cm}}

\vspace{2mm}
\noindent
\small
\end{center}

\vspace*{-3mm}
\begin{center}
{\sf
Figure~1e:~Contribution to the NLO Wilson coefficient $d_{F_L,g}^{(1)}$.
All other conditions
are as in Figure~1a.}
\end{center}
\normalsize

{\footnotesize
\begin{center}
\renewcommand{\arraystretch}{1.5}
\begin{tabular}{||l||r|r|r|r|r|r||} 
\hline \hline
\multicolumn{1}{||c||}{ } &
\multicolumn{6}{|c||}{ max. absolute errors of the {\tt
MINIMAX}-polynomials 
} \\
\hline \hline
\multicolumn{1}{||c||}{Wilson Coeff.} & 
\multicolumn{1}{|c}{$~~\kappa~~$}&
\multicolumn{1}{|c}{$~~\xi = 1~~$}&
\multicolumn{1}{|c}{$~~\xi = 10~~$}& 
\multicolumn{1}{|c}{$~~\xi = 10^2~~$}&
\multicolumn{1}{|c}{$~~\xi = 10^3~~$}&
\multicolumn{1}{|c||}{$~~\xi = 10^4~~$}\\
\hline \hline 
$c_{F_L,g}^{(0)}$  &  0.        & 2.1E-5   & 4.5E-5   & 2.5E-5   & 5.7E-6   
& 2.9E-7        \\
$c_{F_2,g}^{(0)}$  &  0.5        & 1.4E-1   &  8.3E-3  & 3.0E-3   &  1.0E-3  & 3.8E-4       
\\
$c_{g_1,g}^{(0)}$  &  0.        & 1.1E-3   & 6.7E-4   & 2.4E-4   & 8.3E-5   & 3.0E-5       
\\
\hline
$c_{F_L,g}^{(1)}$  &  0.        & 4.1E-5   & 5.0E-5   & 1.4E-5   & 8.9E-6   & 6.9E-7        
\\
$\overline{c}_{F_L,g}^{(1)}$ & 0.  & 2.3E-5   & 5.0E-5   & 1.2E-6   & 1.8E-6   & 1.5E-7    
\\
$c_{F_L,q}^{(1)}$  &  0.        &  1.4E-5  & 2.2E-5   & 4.3E-6   & 3.3E-7   & 4.4E-7        
\\
$\overline{c}_{F_L,q}^{(1)}$ & 0.   & 6.0E-7   & 2.1E-6   & 3.7E-7   & 3.7E-8   & 3.1E-8    
\\
$d_{F_L,q}^{(1)}$ &   0.        &  4.0E-6  & 2.6E-6   & 6.1E-7   & 1.5E-6   & 6.3E-7       
\\ \hline
$c_{F_2,g}^{(1)}$  & 0.4         & 5.6E-2   & 2.6E-2   &  3.9E-3  & 1.0E-3   & 8.5E-4        
\\
$\overline{c}_{F_2,g}^{(1)}$ & 0.   &  8.9E-4  &  5.3E-3  & 1.9E-3   & 6.6E-4   &  2.3E-4   
\\
$c_{F_2,q}^{(1)}$ &  -0.5          &  2.6E-3  & 1.2E-3   & 2.2E-4   & 2.2E-5   & 6.3E-6        
\\
$\overline{c}_{F_L,q}^{(1)}$ & 0.   & 3.2E-4   & 1.3E-4   & 2.2E-5   &  1.8E-6  & 7.1E-7    
\\
$d_{F_2,q}^{(1)}$ &  0.         & 1.3E-4   & 5.1E-5   & 8.1E-6   & 1.0E-4   & 5.8E-4        
\\
$\overline{d}_{F_2,q}^{(1)}$ &  0.  & 1.4E-14   &  8.5E-55  &  --  &  --  & --    \\
\hline\hline
\end{tabular}
\renewcommand{\arraystretch}{1.0}
\end{center}
\vspace{2mm}
\noindent
}
\normalsize
{\sf Table~1: Maximal absolute errors of the {\tt MINIMAX}--polynomials
for the LO and NLO Wilson coefficients as a function of $\xi$.}


We illustrate the principal
response of the Wilson coefficients for the heavy flavor structure functions by the 
following scale--independent
`effective' distributions~: 
\begin{eqnarray}
\label{par1}
f_g(z) &=& 1.8 z^{-0.2}(1-z)^5 \\
f_q(z) &=& 0.6 z^{-0.2}(1-z)^5 \\
g_g(z) &=& 20 z^{1.4}(1-z)^6 \\
\label{par4}
\sum_{q,\overline{q}}^3 e_q^2 f_q(z)&=& 0.15
z^{-0.2}(1-z)^5~. 
\end{eqnarray}
for different structure functions for charm quark production choosing
$m_c = 1.5 \GeV$. The strong coupling constant is fixed to a value
of $\alpha_s = 0.2$ for simplicity. 
All convolutions are performed in Mellin--$N$ space. Finally the structure
functions in $x$--space are obtained by the contour integral
\begin{eqnarray}
\label{eqCONT}
F_i(x) = \frac{1}{\pi} \int_0^\infty dz {\sf Im} \left[ e^{i\Phi}
x^{-c(z)} F(c(z))\right],~~c(z) = c_0 + z e^{i\Phi},~~~\Phi \simeq (3/4) \pi~.
\end{eqnarray}
The absolute error of the {\tt MINIMAX}--representation
if compared to the numerical representation using the
$x$--space parameterizations~\cite{RSN} and the sample distributions
(\ref{par1}--\ref{par4}) in Figures~2a--e below are 
\begin{eqnarray}
\label{eqACCf}
\Delta F_i(x,Q^2) = |F_i^{\tt MINIMAX}(x,Q^2) -
F_i(x,Q^2)|~.
\end{eqnarray}
As seen in the figures below the absolute errors $\Delta F_i(x,Q^2)$ are
mostly smaller than $10^{-5}$ and reach $10^{-4}$ for $g_1(x,Q^2)$ in a 
range of $x < 10^{-2}$.

\begin{center}

\mbox{\epsfig{file=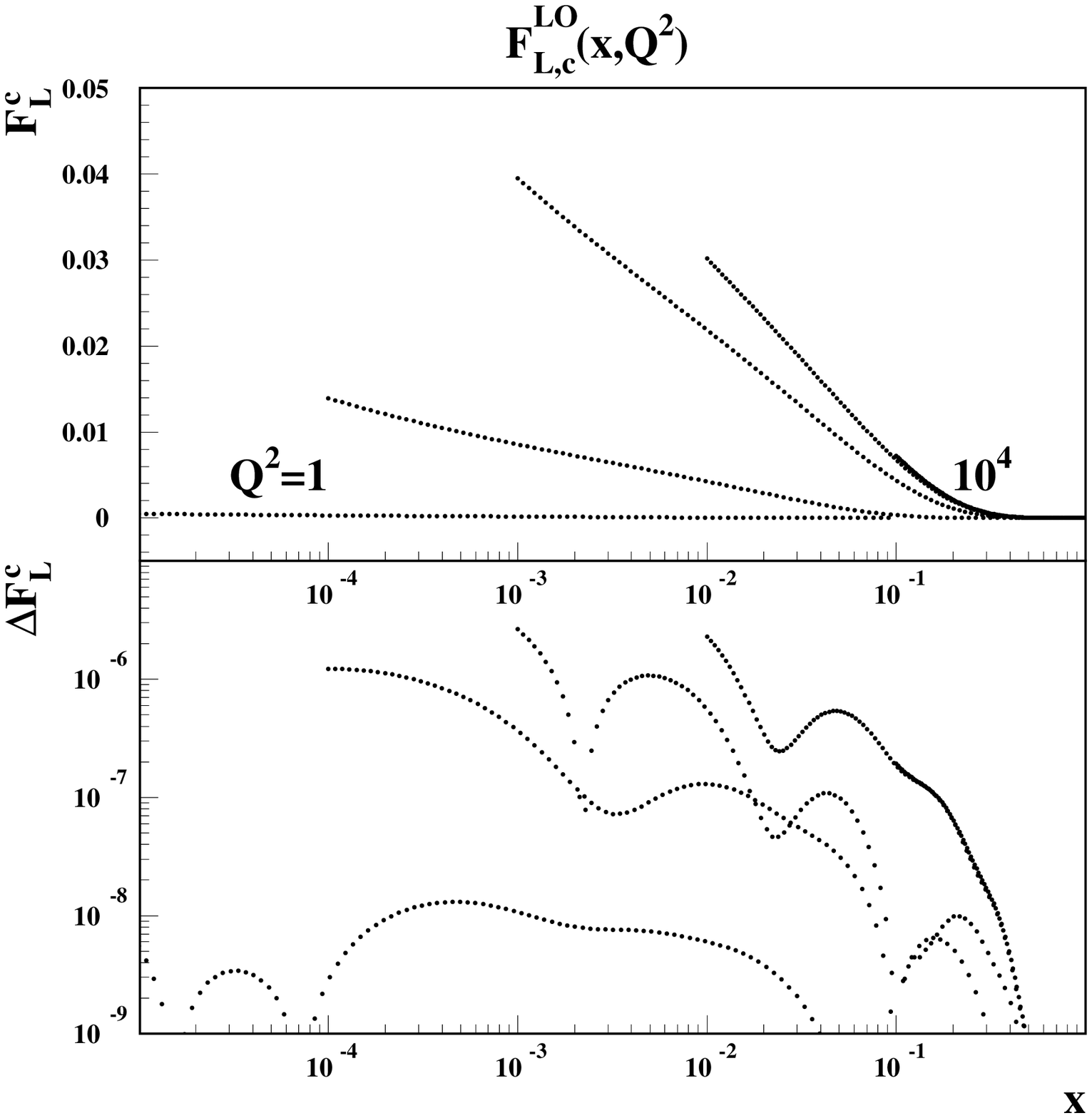,height=10cm,width=10cm}}

\vspace{2mm}
\noindent
\small
\end{center}

\vspace*{-5mm}
\noindent
{\sf
Figure~2a:~$F_{L,c}^{\rm LO}$ in dependence of $x$ and $Q^2$ and the
absolute accuracy of the {\tt MINIMAX}--representation $\Delta F_L^c$.
}
\normalsize

\vspace*{-4mm}
\begin{center}

\mbox{\epsfig{file=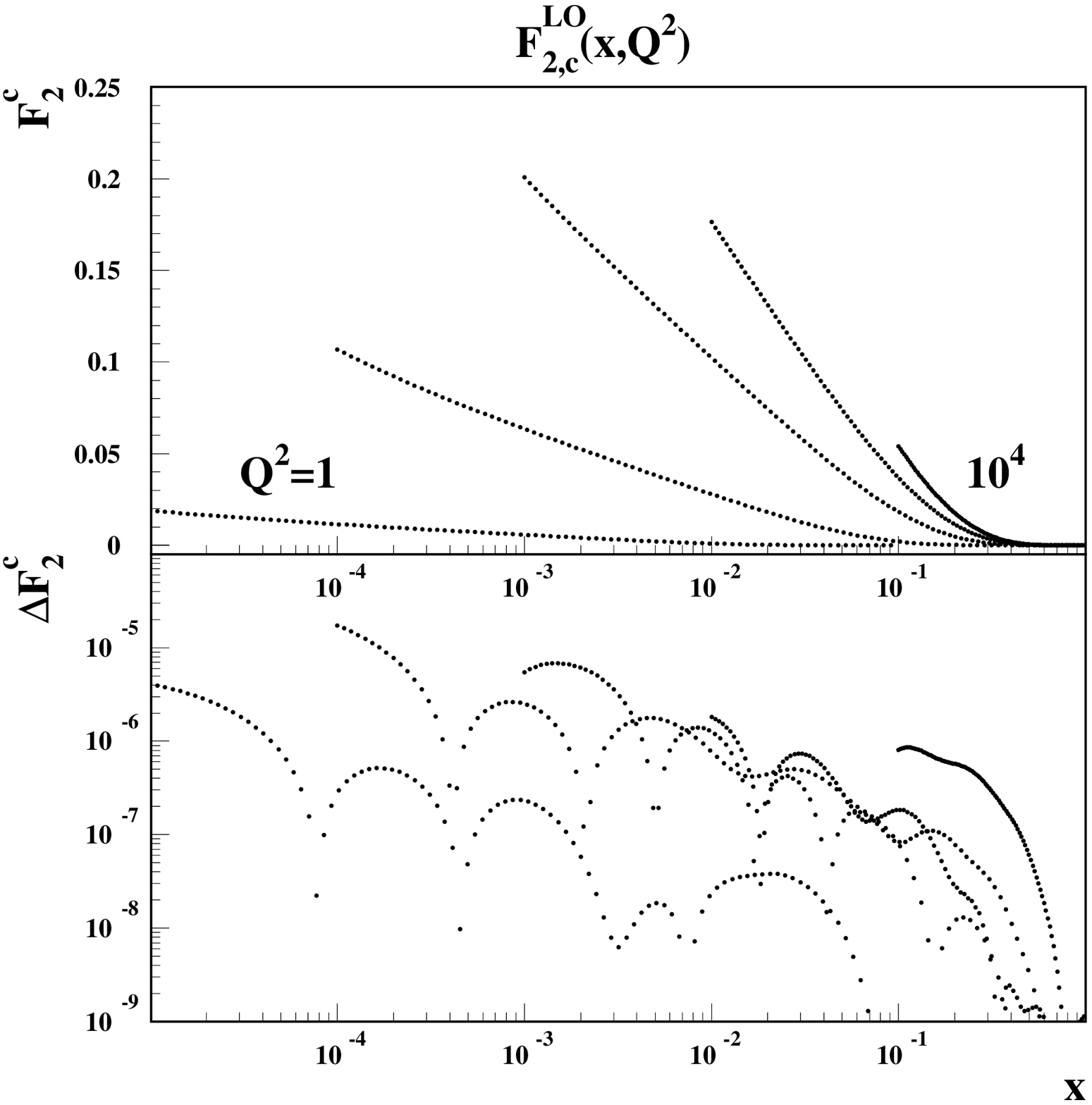,height=10cm,width=10cm}}

\vspace{2mm}
\noindent
\small
\end{center}
{\sf
Figure~2b:~$F_{2,c}^{\rm LO}$ in dependence of $x$ and
$Q^2$ and the
absolute accuracy of the {\tt MINIMAX}--representation $\Delta 
F_2^c$.
}
\normalsize
\begin{center}

\mbox{\epsfig{file=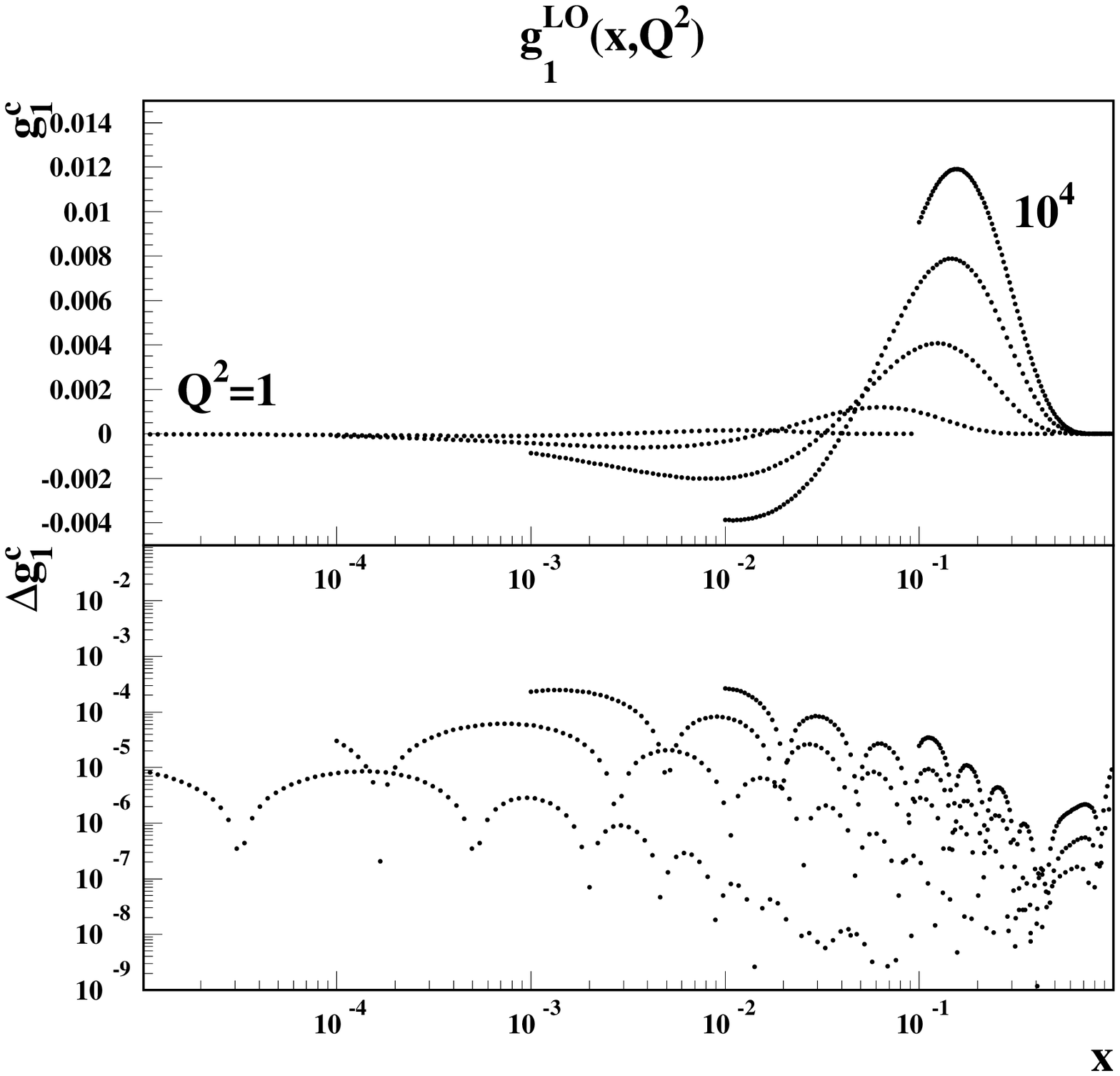,height=10cm,width=10cm}}

\vspace{2mm}
\noindent
\small
\end{center}
{\sf
Figure~2c:~$g_{1,c}^{\rm LO}$ in dependence of $x$ and $Q^2$ and the
absolute accuracy of the {\tt MINIMAX}--representation $\Delta g_1^c$.
}

\normalsize
\begin{center}

\mbox{\epsfig{file=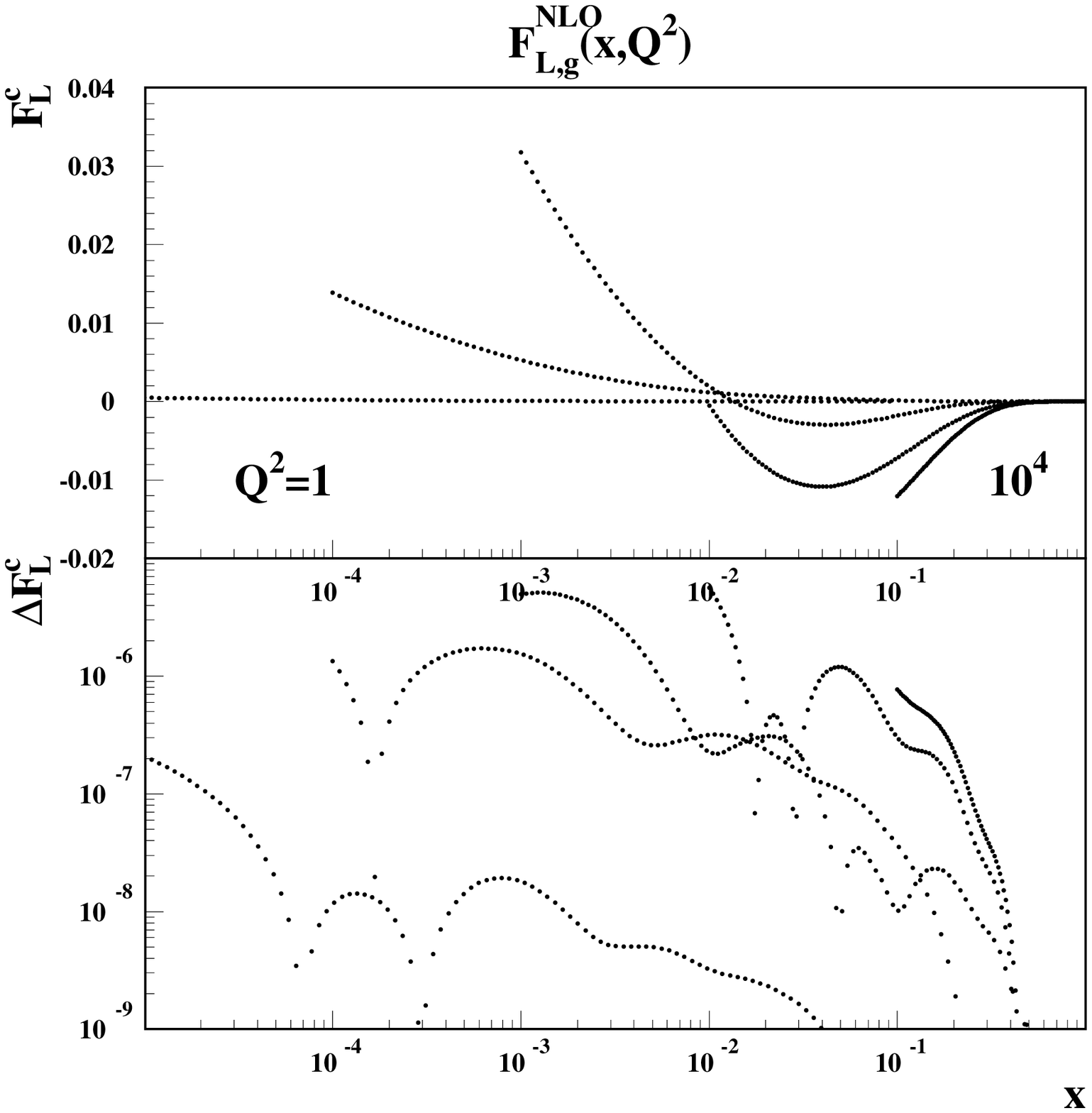,height=10cm,width=10cm}}

\vspace{2mm}
\noindent
\small
\end{center}
{\sf Figure~2d:~Contribution due to $c_{L,g}^{(1)}(\xi,\eta)$ to
$F_{L,c}^{\rm NLO}$ in dependence of $x$ and $Q^2$ and the
absolute accuracy of the {\tt MINIMAX}--representation $\Delta F_L^c$.}
\normalsize
\begin{center}

\mbox{\epsfig{file=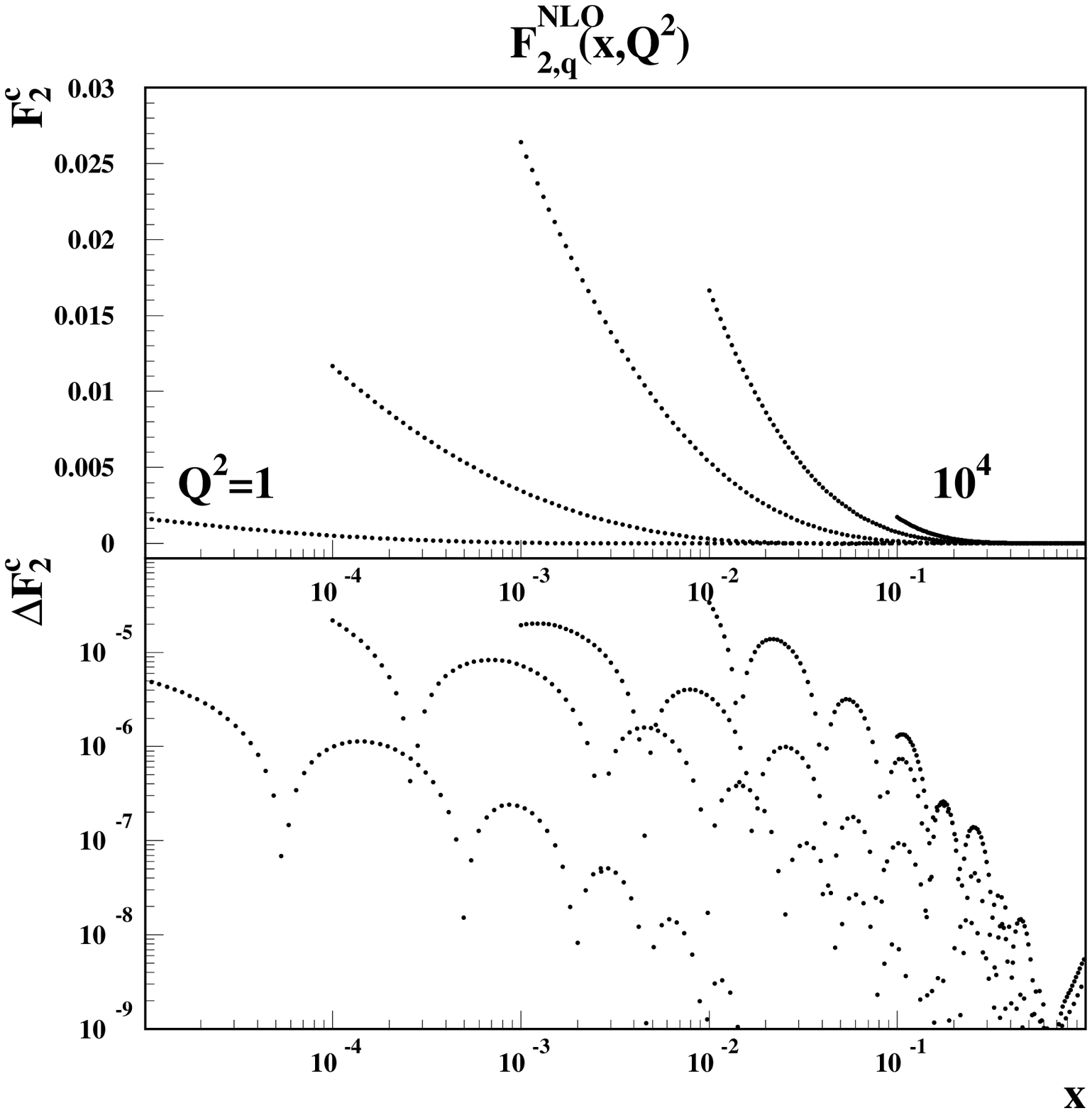,height=10cm,width=10cm}}

\vspace{2mm}
\noindent
\small
\end{center}
{\sf Figure~2e:~Contribution due to $c_{2,q}^{(1)}(\xi,\eta)$ to
$F_{2,c}^{\rm NLO}$ in dependence of $x$ and $Q^2$ and the
absolute accuracy of the {\tt MINIMAX}--representation $\Delta F_2^c$.}
\normalsize

\section{Conclusion}
\label{sec:conc}

\vspace{2mm}
\noindent
We derived semi--analytic representations for the {\sc Mellin} transforms
of the Wilson coefficients for heavy flavor production in unpolarized and 
polarized deeply inelastic scattering to two--loop order for complex 
values of $N$. These representations are obtained using Riemann--Liouville
fractional integrals and are meromorphic functions. The expansion 
coefficients were determined using the {\tt MINIMAX}--method at fixed 
values of $\xi = Q^2/m^2$, which allows to obtain very precise 
approximations. This representation of the one-- and two--loop massive
Wilson coefficients is used in QCD evolution programs  solving the 
renormalization group equations for mass factorization in {\sc 
Mellin}--space. These evolution programs are particularly fast due to the 
fact that all coefficient functions and anomalous dimensions are only 
once calculated during the initialization of the code. In the case of the 
heavy flavor coefficient functions the whole array in $\xi$ is stored for 
the inversion contour as a grid allowing  for precise and fast numerical 
interpolations during the fit of the non--perturbative parton distribution 
functions. This representation of the Wilson coefficients can be directly 
used to implement the heavy--flavor effects into the physical evolution 
kernels for factorization scheme--invariant singlet evolution.

\vspace{1mm} \noindent
The parameterization can be obtained from 
{\tt www-zeuthen.desy.de/theory/research/
num.html}

\vspace{2mm} 
\noindent 
{\bf Acknowledgment.} This
paper was supported in part by DFG Sonderforschungsbereich Transregio~9,
Computergest\"utzte Theoretische Physik.

\newpage
\clearpage
\portrait


\begin{thebibliography}{99}
%
\bibitem{XSPAC}
L.F. Abbott, W.B. Atwood, and R.M. Barnett, Phys. Rev. {\bf D22} (1980)
582;\\
K. Kato, Z. Shimizu, and H. Yamamoto, Progr. Theor. Phys. {\bf 63} (1980)
1295;\\ 
K. Kato and Z. Shimizu, Progr. Theor. Phys. {\bf 64} (1980) 703;\\
A. Devoto, D.W. Duke, J.F. Owens, and R.G. Roberts, Phys. Rev. {\bf D27}
(1983) 508;\\
M. Virchaux and A. Oraou, Saclay preprint DPhPE 87--15;\\
J.F. Botts et al., Phys. Lett. {\bf B304} (1993) 159;\\
C. Pascaud and F. Zomer, H1 Note, {\tt H1-11/94-404};\\
M. Miyama and S. Kumano, Comput. Phys. Commun. {\bf 94} (1996) 185;\\
M. Botje, {\tt hep-ph/9707289};\\                  
S.I. Alekhin, Eur. Phys. J. {\bf C10} (1999) 395;\\ 
M. Hirai, S. Kumano and M. Miyama, Comput. Phys. Commun. {\bf 108} (1998)
38;\\ 
A. Cafarella and  C. Coriano, {\tt hep-ph/0311313}. 
%
\bibitem{ORTHPOL}
F.J. Yndurain, Phys. Lett. {\bf B74} (1978) 68;\\
G. Parisi and N. Sourlas, Nucl. Phys. {\bf B151} (1979) 421;\\
W. Furmanski and R. Petronzio, Nucl. Phys. {\bf B195} (1982) 237;\\
J. Chyla and J. Rames, Czech. J. Phys. {\bf B36} (1986) 567; Z. Phys. {\bf
C31} (1986) 151;\\
V.G. Krivokhizin et al., Z. Phys. {\bf C36} (1987) 51;\\
J. Bl\"umlein, M. Klein, G. Ingelman, and R. R\"uckl,
Z. Phys. {\bf C45} (1990) 501;\\ 
S. Kumano and J.T. Londergan, Comput. Phys. Commun. {\bf 69} (1992) 373;\\ 
R. Kobayashi, M. Konuma, S. Kumano, Comput. Phys. Commun.
{\bf 86} (1995) 264;\\
G. Shaw, Nucl. Phys. {\bf A675} (2000) 84c.
%
\bibitem{MSPACE}
M. Gl\"uck and E. Reya, Phys. Rev. {\bf D14} (1976) 3034;\\
M. Diemoz, F. Ferroni, E. Longo, and G. Martinelli, Z. Phys. {\bf C39}
(1988) 21;\\
M. Gl\"uck, E. Reya, and A. Vogt, Z. Phys. {\bf C53} (1992) 127;\\
M. Gl\"uck, E. Reya, M. Stratmann, and W. Vogelsang,
Phys. Rev. {\bf D53} (1996) 4775;\\ 
R.K. Ellis, Z. Kunszt and E.M. Levin 
Nucl. Phys. {\bf B420} (1994) 517; E: {\bf B433} (1995) 498:\\  
R.D. Ball and S. Forte, (CERN),
Phys. Lett. {\bf B358} (1995) 365;
{\bf B378} (1996) 255.\\
D. Kosower, Nucl. Phys. {\bf B506} (1997) 439;\\ 
S. Forte, L. Magnea, A. Piccione, and G. Ridolfi,
Nucl. Phys. {\bf B594} (2001) 46;\\
S. Weinzierl, Comput. Phys. Commun. {\bf 148} (2002) 312. 
%
\bibitem{BB}
J. Bl\"umlein and H. B\"ottcher, Nucl. Phys. {\bf B636} (2002) 225.
%
\bibitem{COMP}
J. Bl\"umlein et al., 
{\sf A detailed comparison of NLO QCD evolution codes}, {\tt
hep-ph/9609400};\\ 
W. Giele et al., {\sf The QCD/SM working group: summary report},
{\tt hep-ph/0204316}. 
%
\bibitem{SX}
J.~Bl\"umlein and A.~Vogt, 
Phys.\ Lett.\  {\bf B370} (1996) 149; {\bf B386} (1996) 350;
Acta Phys. Polon. {\bf B27} (1996) 1309;
Phys.\ Rev.\  {\bf D57} (1998) 1; 
{\bf D58} (1998) 014020;\\
J.~Bl\"umlein, S.~Riemersma and A.~Vogt,
Nucl.\ Phys.\ Proc.\ Suppl.\  {\bf 51C} (1996) 30;
Eur. Phys. J. {\bf C1} (1998) 255;\\
J. Bl\"umlein, V. Ravindran, W.L. van Neerven, and A. Vogt,
{\tt hep-ph/9806368};\\
R.D. Ball and S. Forte, {\tt hep-ph/9805315};\\ 
J. Bl\"umlein, {\tt hep-ph/9909449};\\
J. Bl\"umlein and H. Kawamura, Acta Phys. Polon. {\bf B33}
(2002) 3719.
%
\bibitem{SI}
W. Furmanski and R. Petronzio, Z. Phys. {\bf C11} (1982) 293;\\
G. Grunberg, Phys. Rev. {\bf D29} (1984) 2315;\\
S. Catani, Z. Phys. {\bf C75} (1997) 665;\\
J. Bl\"umlein and A. Vogt (1999), unpublished.
%
\bibitem{BRN1}
J. Bl\"umlein, V. Ravindran, and W.L. van Neerven, Nucl. Phys. {\bf B586}
(2000) 349.
%
\bibitem{BK}
L. Baulieu and C. Kounnas, Nucl. Phys. {\bf B155} (1979) 429;\\
M. Gl\"uck, E. Hoffmann, and E. Reya, Z. Phys. {\bf C13} (1982)  119.
%
\bibitem{LOUNP}
E. Witten, Nucl. Phys. {\bf B104} (1976) 445;\\
J.P. Leveille and T. Weiler, Nucl. Phys. {\bf B147} (1979) 147.
%
\bibitem{SVZ}
M.A. Shifman, A.I. Vainshtein, and V.I. Zakharov,
Nucl. Phys. {\bf B136} (1978) 157.
%
\bibitem{LOPOL}
A.D. Watson, Z. Phys. {\bf C12} (1982) 123;\\
W. Vogelsang, Z. Phys. {\bf C50} (1991) 275.
%
\bibitem{HFNLO}
E. Laenen, S. Riemersma, J. Smith, and W.L. van Neerven,
Nucl. Phys. {\bf B392} (1993) 162, 229. 
%
\bibitem{RSN}
S. Riemersma, J. Smith, and W.L. van Neerven, Phys. Lett. {\bf B347}
(1995) 143; numerical update: S. Riemersma, 1996.
%
\bibitem{BNR2}
J. Bl\"umlein, V. Ravindran, and W.L. van Neerven, Phys. Rev. 
{\bf D68} (2003) 114004.
%
\bibitem{MINIM1}
C. Hastings, jr., {\sf Approximations for Digital Computers}, (Princeton
University Press, Princeton/NJ, 1953);\\
M. Abramowitz and I.A. Stegun, {\sf Handbook of Mathematical Functions},
(NBS, Washington, 1964);\\
L.A. Lyusternik, O.A. Chervonenkis, and A.R. Yanpol'skii,
{\sf Handbook for Computing of Elementary Functions}, Russian ed.
(Fizmatgiz, Moscow, 1963); (Pergamon Press, New York, 1964).
%
\bibitem{MINIMAX}
{\tt Maple V}, Release 5, Waterloo Maple Inc.;\\
{\tt  mathematica 3.0}, Wolfram Research Inc.;\\ NAG - Numerical 
Algorithms
Group, {\tt http://www.nag.co.uk/}.
%
\bibitem{ANCONT}
J. Bl\"umlein, Comp. Phys. Commun. {\bf 133} (2000) 76.
%
\bibitem{HO1}
J. Bl\"umlein and S. Kurth, DESY 97-160, {\tt hep-ph/9708388};
Phys. Rev. {\bf D60} (1999) 014018;\\
J. Bl\"umlein, {\sf Algebraic Relations Between Harmonic Sums and
Associated  Quantities}, DESY 03--134, {\tt hep-ph/0311046}, in print.
%
\bibitem{BAT}
A. Erd\'elyi, W. Magnus, F. Oberhettinger and  F.G. Tricomi, {\sf
The Bateman Manuscript Project:~Tables of Integral Transforms},
{\bf Vol. II} (McGraw--Hill, New York, 1954), pp.~181.
\end{thebibliography}
\end{document}